\documentclass[conference,english]{IEEEtran}[2015/08/26]

\usepackage{iftex}
\usepackage{upquote}

\usepackage{amssymb}
\usepackage{amsfonts}
\usepackage[ngerman,main=english]{babel}
\addto\extrasenglish{\languageshorthands{ngerman}\useshorthands{"}}
\ifluatex
	\usepackage[no-math]{fontspec}
	\usepackage{unicode-math}
	\usepackage{selnolig}
\else
	\usepackage[zerostyle=b,scaled=.75]{newtxtt}
	\usepackage[T1]{fontenc}
\fi
\usepackage[
	babel=true, % Enable language-specific kerning. Take language-settings from the languge of the current document (see Section 6 of microtype.pdf)
	expansion=alltext,
	protrusion=alltext-nott, % Ensure that at listings, there is no change at the margin of the listing
	final % Always enable microtype, even if in draft mode. This helps finding bad boxes quickly.
]{microtype}
\DisableLigatures{encoding = T1, family = tt* }
\usepackage{graphicx}
\usepackage{diagbox}
\ifluatex
	\usepackage{spelling}
	\spellingoutput{off}
\fi

\usepackage[dvipsnames, table]{xcolor}
\usepackage{listings}

\definecolor{lstbg}{gray}{0.97}
\definecolor{lstframe}{gray}{0.55}
\definecolor{lsttitlebg}{gray}{0.20}

\definecolor{eclipseStrings}{RGB}{42,0.0,255}
\definecolor{eclipseKeywords}{RGB}{127,0,85}
\colorlet{numb}{magenta!60!black}
\ifpdftex
\fi

\usepackage[autostyle=true]{csquotes}
\defineshorthand{"`}{\openautoquote}
\defineshorthand{"'}{\closeautoquote}
\usepackage{booktabs}
\usepackage{paralist} % Extended enumerate, such as \begin{compactenum}

\usepackage[%
	square,        % for square brackets
	comma,         % use commas as separators
	numbers,       % for numerical citations;
	sort&compress  % as sort but in addition multiple numerical citations are compressed if possible (as 3-6, 15);
]{natbib}

\usepackage{etoolbox}
\makeatletter
\patchcmd{\NAT@test}{\else \NAT@nm}{\else \NAT@hyper@{\NAT@nm}}{}{}
\makeatother

\usepackage{multirow} % Mehrfachspalten
\usepackage{dcolumn}  % Ausrichtung an Komma oder Punkt

\usepackage{stfloats}
\fnbelowfloat

\newcommand{\sidecomment}[1]{}

\makeatletter
\@ifundefined{missingfigure}{}{}
\@ifundefined{textcomment}{}{}
\@ifundefined{sidecomment}{\newcommand{\sidecomment}[1]{\marginpar{#1}}}{}
\@ifundefined{todo}{}{}
\@ifundefined{TODO}{}{}
\@ifundefined{todofix}{}{}
\@ifundefined{change}{}{}
\makeatother

\usepackage[group-minimum-digits=4,per-mode=fraction]{siunitx}

\usepackage[font=footnotesize]{caption}
\usepackage[incolumn]{mindflow}

\usepackage[capitalise,nameinlink,noabbrev]{cleveref}
\crefname{listing}{Listing}{Listings}
\Crefname{listing}{Listing}{Listings}
\crefname{lstlisting}{Listing}{Listings}
\Crefname{lstlisting}{Listing}{Listings}
\usepackage{lipsum}
\usepackage[math]{blindtext}
\usepackage{mwe}
\usepackage[realmainfile]{currfile}
\usepackage{tcolorbox}
\tcbuselibrary{listings}
\DeclareFontFamily{U}{MnSymbolC}{}
\DeclareSymbolFont{MnSyC}{U}{MnSymbolC}{m}{n}
\DeclareFontShape{U}{MnSymbolC}{m}{n}{
	<-6>    MnSymbolC5
	<6-7>   MnSymbolC6
	<7-8>   MnSymbolC7
	<8-9>   MnSymbolC8
	<9-10>  MnSymbolC9
	<10-12> MnSymbolC10
	<12->   MnSymbolC12%
}{}
\DeclareMathSymbol{\powerset}{\mathord}{MnSyC}{180}

\usepackage[
	nolist,
	nolinebreak
]{acronym}

\usepackage{xspace}
\makeatletter
\xspaceaddexceptions{\grqq \grq \csq@qclose@i \} }
\makeatother

\newcommand{\eg}{e.g.,\ }
\newcommand{\ie}{i.e.,\ }

\makeatletter
\newcommand{\hydash}{\penalty\@M-\hskip\z@skip}
\makeatother

\input{ushyphex}

\ifpdftex
	\input{glyphtounicode}
\fi

\begin{acronym}
  \acro{API}{application programming interface}
  \acro{RAN}{radio access network}
  \acro{IBN}{intent-based networking}
  \acro{IBS}{intent-based system}
  \acro{PBC}{policy-based control}
  \acro{LLM}{large language model}
  \acro{RAG}{retrieval-augmented generation}
  \acro{MCP}{model context protocol}
  \acro{ISP}{internet service provider}
  \acro{SAIN}{service assurance for intent-based networking}
  \acro{NF}{network function}
  \acro{OA}{orchestrator agent}
  \acro{AIR}{agent information repository}
  \acro{GUI}{graphical user interface}
  \acro{AI}{artificial intelligence}
  \acro{NAT}{network address translation}
\end{acronym}

\begin{document}
\title{Enhancing Secure Intent-Based Networking with an Agentic AI: The EU Project MARE Approach}

\author{%
  \IEEEauthorblockN{Iulisloi Zacarias\IEEEauthorrefmark{1}, Marla Grunewald\IEEEauthorrefmark{1}, Fin Gentzen\IEEEauthorrefmark{1}, Xavi Masip-Bruin\IEEEauthorrefmark{2} Admela Jukan\IEEEauthorrefmark{1}}
  \IEEEauthorblockA{\IEEEauthorrefmark{1}Institut f\"ur Datentechnik und Kommunikationsnetze,  
  Technische Universit\"at Braunschweig, Germany}
  \IEEEauthorblockA{\IEEEauthorrefmark{2}Advanced Network Architectures Lab (CRAAX), Universitat Polytecnica de Catalunya, Spain}
  Email: \{i.zacarias, marla.grunewald, f.gentzen, a.jukan\}@tu-bs.de, xavier.masip@upc.edu
}


\maketitle
% In case you want to add a copyright statement.
% Works only in the compsoc conference mode.
%
% Source: https://tex.stackexchange.com/a/325013/9075
%
% All possible solutions:
%  - https://tex.stackexchange.com/a/325013/9075
%  - https://tex.stackexchange.com/a/279134/9075
%  - https://tex.stackexchange.com/q/279789/9075 (TikZ)
%  - https://tex.stackexchange.com/a/200330/9075 - for non-compsocc papers
\iffalse
  \IEEEoverridecommandlockouts
  \IEEEpubid{\begin{minipage}{\textwidth}\ \\[12pt] \centering
      1551-3203 \copyright 2015 IEEE.
      Personal use is permitted, but republication/redistribution requires IEEE permission.
      \\
      See \url{https://www.ieee.org/publications_standards/publications/rights/index.html} for more information.
    \end{minipage}}
\fi % disabled by default

%%% ===============================================================================
%   For Submission EDAS
%   https://edas.info/N34224
%   Track: NET – Network Softwarisation
%   Keywords:
%    - intent-based networking
%    - multi-agent
%    - network management
%    - multi-domain
%%% ===============================================================================

%%% ===============================================================================
%%% Sections
%%% ===============================================================================
\begin{abstract}
In the EU project MARE, a novel plane was proposed, and used in combination with \ac{IBN}, allowing the operator to focus on \emph{what}, rather than on \emph{how}.  Recently, \acp{LLM} have been successfully employed to translate the high-level intents into low-level actions. The open challenge is to understand how \ac{IBN} can be effectively enhanced with \ac{LLM} and the emerging \emph{agentic AI} for security purposes. Enhancing \ac{IBN} with an agentic AI paradigm introduces significant challenges that existing solutions do not fully address. This paper proposes an enhanced \ac{IBN} framework with a strong security focus  toward agentic AI. We address the architectural and security requirements for a multi-agent \ac{IBS} architecture, including a multi-domain \ac{IBN}. We propose a hierarchical multi-agent and multi-vendor architecture, which can also be applied more broadly in 6G architectures and beyond the security architecture proposed in MARE. The architecture incorporates an interactive intent-processing pipeline using \acp{LLM}, and it also allows the \ac{IBS} to connect to external security knowledge bases, such as MITRE ATT\&CK, MITRE FiGHT, and NIST. 
\end{abstract}

% No keywords according to the LaTeX model
\acresetall
\section{Introduction}

Next generation mobile networks (6G and beyond) are characterized by the unprecedented complexity in management of the infrastructure, deeming the  traditional network management approaches obsolete, when relying on static rules, \ac{PBC}, and predefined algorithms. The \ac{IBN}\cite{rfc9315} paradigm aims to address network management complexity while still allowing the network operator to express their needs and goals. In the EU project MARE, a novel plane is proposed, referred to as a Security Plane, akin to data and control planes in networking. The new plane, in combination with \ac{IBN}, allows the operator to specify \textit{what} should be achieved (the intent), rather than \textit{how} the infrastructure should be configured to reach a specific goal. 

Since recently \acp{LLM} have been successfully employed to translate these high-level intents into low-level actions, the open challenge is to understand how IBN can be effectively enhanced with LLM and the emerging \emph{agentic AI} beyond just parsing semi-structured data and simple reasoning about network anomalies~\cite{dzeparoska2024,gu2025}. 

Enhancing \ac{IBN} with an agentic AI paradigm introduces significant challenges that existing solutions do not fully address. Although \ac{LLM} models have shown promise in enhancing IBN toward the full implementation of \acp{IBS}, they introduce new security vulnerabilities, such as malicious intent injection and model inference attacks. Additionally, the requirement of multi-domain interoperability and the need for agentic intelligence to deal with different technologies to handle autonomous operations creates a gap in the current proposed architectures. Finally, what remains unaddressed are the evaluation metrics for network management systems, to provide insights and explainability, at both the overall intent implementation validity, as well as the individual decision points, \eg which management tools to deploy. Understanding how to secure the network against external threats using \ac{IBN} enhanced with agentic AI is a novel and open area of research and development.
The goal of this paper is to propose an \ac{IBN}-based 6G architecture that leverages \acp{LLM} and agentic AI to enhance 6G architecture security capabilities, inherently integrating a novel Security Plane. Specifically, this paper tries to address the following questions:
\begin{compactenum}
  \item What are the architectural and security requirements for a multi-agent \ac{IBS} architecture?
  \item How can a multi-domain, agentic \ac{IBN} framework be designed to support security intents on multi-vendor networks?
  \item How can \acp{LLM} be effectively employed in a multi-agent \ac{IBS}? 
  \item How can external sources of information be used to enhance the capabilities and reasoning of the \ac{IBS}, and protect the network against zero-day threats?
  \item How can we verify that the proposed architecture accurately calls the proper methods and tools as defined by the intent, and which performance metrics we should use to this end?
\end{compactenum}

We address the above questions by proposing a multi-agent architecture for an \ac{IBS} with focus on network security, based on the security plane architecture envisaged by MARE project. We propose a hierarchical multi-agent architecture capable of operating in multiple domains and a multi-vendor network, which can also be applied more broadly. Additionally, the proposed architecture is extensible as it can accommodate third-party agent plugins. The architecture incorporates an interactive intent-processing pipeline using \acp{LLM} and also allows the \ac{IBS} to connect to external security knowledge bases, such as MITRE ATT\&CK, MITRE FiGHT, and NIST, to enhance its intent mapping capabilities.

The remainder of this paper is organized as follows. \Cref{sec:bg-and-rw} reviews the main concepts of \ac{IBN} applied to next-generation mobile communication networks and details the comprehensive requirements for an \ac{IBS} in 6G, covering architecture, security, and telemetry. \Cref{sec:related-work} reviews existing solutions and highlights the gaps in current \ac{LLM}-based and agentic \ac{IBN} approaches. \Cref{sec:architecture} presents our new architecture, detailing the role of the Orchestrator Agent and the intent processing pipeline. Finally, \Cref{sec:verification} presents performance results and discusses the IBS verification. \Cref{sec:conclusions} concludes the paper and provides an outlook.

\section{Background and related work}
\label{sec:bg-and-rw}
\subsection{IBN Systems in 6G Networks}
\label{sec:bg-ibn}

At the architectural level, an \ac{IBS} for 6G must support an end-to-end closed-loop workflow comprising intent profiling, translation, resolution, activation, and assurance, as described in \cite{leivadeas2022survey}. The architecture must expose a logically centralized but physically distributed intent interface that can be consumed by network operators, vertical industries, and higher-level orchestration frameworks. This interface should allow intents to be expressed in natural or controlled languages. At the same time, intents should be mapped internally to abstract service models, such as network services and slices, defined using predefined data structures.

At the functional level, the \ac{IBS} must maintain a clear separation of goals between the intent layer and the underlying control and data planes. \ac{IBN} must operate above  any system controller, such as SDN Controllers and RAN Intelligent Controllers, having the responsibility of interpreting intents, deriving high-level policies, and orchestrating their deployment on heterogeneous network functions and devices \cite{leivadeas2022survey, SecurityIBNAhmad2023}.    In 6G networks, it is important to explicitly align with 3GPP service-based architectures and network functions such as AMF, SMF, PCF, and NWDAF, as well as their exposure via the Network Exposure Function (NEF) and external interfaces \cite{Ahn2025I2NSF5G}. The I2NSF-based architecture for 5G edge security demonstrates that an \ac{IBS} must be able to generate both network-level and application-level policies and deliver them to core components and user equipment,  while supporting mobility scenarios and dynamic session migration.

Another functional requirement to be considered is the human interaction  between the system operator and the \ac{IBS}. The interfaces must support role differentiation, feedback mechanisms, and abstraction levels that accommodate both telecom experts and non-expert operators \cite{tu2025intent}. Therefore, the \ac{IBN} architecture must integrate solutions and  mechanisms for intent validation and explanation of translation output.

Human-in-the-loop and policy-aligned operation are essential for trustworthy deployment of agentic AI components, as mentioned in \cite{zhang2026toward}. This means that the \ac{IBS} must support configurable human intervention points, rollback mechanisms for erroneous configurations, and comprehensive observability of both network state and decision processes.

Standardization efforts by ONF, ETSI ENI, 3GPP, ITU-T, and the IETF SAIN working group demonstrate that interoperable intent models, assurance frameworks, and management interfaces are essential to prevent vendor lock-in \cite{Claise2023SAIN}. An \ac{IBS} targeting 6G must align with these initiatives by adopting common data models (e.g., YANG-based network services), open APIs for assurance and telemetry, and semantics for expressing performance, security, and reliability intents that can be understood across domains, which is an open challenge.

Furthermore, \acp{IBS} should be vendor agnostic, being able to receive the intent from the operator, process it, and decompose it in commands and policies that will work in the network, independently of the devices' model or vendor.
%\hl{done}

\subsection{Security and Privacy Requirements}
Security is one of the most critical requirements for 6G-capable \ac{IBS}. At the intent acquisition stage, threats such as malicious intent injection and mistakes by under-skilled personnel must be mitigated through strong authentication, secure authorization, and accountability mechanisms that precisely map intents to verified identities \cite{SecurityIBNAhmad2023}. The human-friendly nature of \ac{IBN} interfaces must not compromise access control. Thus,  the ease-of-use and security must be jointly addressed, considering the requirement of a secure credential management, multi-factor authentication, and auditable logging of all intent operations.

The translation and conflict resolution phases introduce additional attack possibilities. Complex or ambiguous intents can be exploited to create semantic loopholes, misconfigurations, or unseen policy conflicts.  The work in \cite{SecurityIBNAhmad2023} proposed that coarse-grained control and AI-assisted interpretation can lead to inconsistent or faulty results, particularly when large language models are used to parse natural-language intents. This behavior forces a solution designer to use a  deterministic or constrained intent language, formal verification or testing of generated configurations, and conflict-resolution mechanisms that consider both functional and security policies across multiple tenants and domains.
Furthermore, \ac{IBN} inherits the security issues that are faced in %vulnerabilities of 
SDN-centric architectures, including controller-targeted denial of service (DoS/DDoS) attacks and software exploits in centralized controllers \cite{Pillai2024SDNDDoS,SecurityIBNAhmad2023}. Therefore, 6G-oriented \acp{IBS} must  deploy resilient, distributed control-plane designs, including replicated controllers  and strict resource isolation for \ac{IBN} and SDN components. Mechanisms for rate-limiting, anomaly detection, and automated isolation of malicious flows at the data plane are necessary to protect intent controllers from volumetric and low-rate DDoS attacks. 

In the context of privacy,  it is required that intents and associated policy attributes do not leak sensitive business logic or user information.  Multi-domain \ac{IBN} coordination, as studied in \cite{SecurityIBNAhmad2023}, proposes a confidentiality-preserving Northbound Interfaces (NBIs) and authorization schemes that prevent unauthorized domains from accessing intent semantics beyond what is needed for service realization. Using AI/ML models for intent processing introduces new vulnerabilities that need to be considered, such as model stealing and inference attacks. Protection against those attacks is necessary to prevent adversaries from recovering training data or internal model behavior. Applications of LLM decision-making per model can furthermore protect the privacy of domain-specific data by sharing the decision-making results rather than data.

\subsection{Telemetry, Data Models, and Assurance}

The goal of service assurance for \ac{IBN} is to build a system that can verify, at runtime, whether services instantiated from intents are operating as expected. RFC 9417~\cite{Claise2023SAIN}  defines an architecture implementing a \ac{SAIN}. This  architecture  constructs an assurance graph mapping services to components called subservices, allowing operators to correlate service degradation with specific network  root cause/symptoms. This information is used to alert the operational team where to focus its attention for maximum return, by focusing its priority on the degrading/failing components impacting the highest number of its customers, especially customers with the Service-Level Agreement (SLA) contracts involving penalties in case of failure \cite{Claise2023SAIN}. An \ac{IBS} for 6G must therefore integrate model-driven telemetry, assurance graphs, and expression graphs to enable such correlation at scale. Model-driven telemetry based on YANG data models, as explained in \ac{SAIN}~\cite{Claise2023SAIN}, is crucial to avoid inconsistent monitoring across management protocols and data sources. By reusing the same data models for configuration and telemetry, the \ac{IBS} can maintain a coherent view of service state and use that in its closed-loop assurance logic.

The I2NSF-based 5G edge security framework in  \cite{Ahn2025I2NSF5G} further illustrates that continuous monitoring and analytics are required not only at the core but also at the edge, using Security Data Analytics Functions (SDAFs) and monitoring storage that aggregate reports from distributed enforcement points. Therefore, \acp{IBS} must provide open, standardized interfaces for telemetry ingestion and expose assurance results to higher-level orchestration layers in a machine-readable way.
%\hl{done}

\subsection{Intent Modeling and Data/Knowledge}
Intent Modeling is a crucial phase to ensure the quality of \ac{IBN} behavior. Translating low-level configurations such as \ac{NAT} rules into high-level, vendor-independent intent languages allows the network administrators and operators to focus on policies and migrate legacy configurations into \acp{IBS}~\cite{ribeiro2022deterministic}. Thus,  6G-oriented \acp{IBS}  must support both top-down and bottom-up intent workflows: operators express new intents, while the system can also infer and reconstruct intents from deployed configurations to maintain consistency.

In the context of data and knowledge, 6G deployments will rely on large repositories of intents mapped into configurations, performance outcomes, and assurance results. These repositories can serve as training datasets for LLM-based intent translation systems, similar to~\cite{tu2025intent}, where fine-tuning, dynamic in-context learning, and continuous learning are applied to improve translation accuracy while maintaining reasonable model sizes and processing times. To this end, an \ac{IBS} must  store and maintain datasets of historical intents, configurations, and outcomes, while assuring its privacy and anonymity. 

\subsection{AI, Learning, and Agentic-AI Requirements}

The increasing integration of \ac{LLM}-based agentic \ac{AI} introduces new requirements to autonomous network management systems. \cite{zhang2026toward} defines four design principles for such systems: compactness, efficiency, knowledge and reasoning, and migration. Those principles are directly applicable to 6G \ac{IBN}: models must be lightweight enough to run at the edge, efficient enough to meet latency constraints, capable of explicit reasoning over multi-modal state, and able to transfer knowledge across tasks and domains. \ac{IBN} architectures must therefore support model placement strategies across the cloud and edge, as well as dynamic offloading for inference tasks. Additionally, the system should implement memory and retrieval mechanisms to allow agents to access historical intents, policies, and assurance results. \ac{IBS} architectures proposed in \cite{antonopoulos2025agile} and \cite{tong2025core} aim to incorporate these principles by using role-based interfaces, shared ontologies/databases, and policy-aligned operations, including human-in-the-loop supervision for sensitive decisions. 

Adaptability of ML models is also required in an Agentic AI system, as mentioned in \cite{xiao2025}, where existing AI network solutions are built on closed-loop, passive learning. Models are trained on static data, and the tool creators assume the same data will be present in the network. Since the device's presence in the network is ephemeral, topology changes and new threats (like 0-day threats) pose challenges to those models, which may present unseen scenarios. Dealing with this type of thread requires moving from a model that simply ``knows'' a fixed dataset to one that can ``reason'' through a scenario it has never seen before, using the most up-to-date thread and mitigation information.

\subsection{Surveying and categorizing agentic IBN Approaches}
\label{sec:related-work}

In this section, we survey and categorize existing \ac{IBN} frameworks, focusing on analyzing \acp{IBS} that adopt an agentic AI paradigm, while taking the security aspect into account. %%%%%%%%%%%%% working here
We are particularly interested in whether attacks on the infrastructure are explicitly addressed, and if so, how they are handled. We also examine whether the framework can be used for multiple domains. Furthermore, we analyze the agent communication flow. 
Based on the survey shown in \cref{tab:comparison}, we divided the examined frameworks into two groups. Our categorization focused primarily on Secure \ac{IBN} and Agentic workflow topology, as we could not find any other frameworks that concentrate exclusively on security in 6G networks. \par
The first group, consisting of papers \cite{chatzistefanidis2026}, \cite{Brodimas20257150}, \cite{Chatzistefanidis2024227} focuses each on specific security aspects. Although none of the three articles presented focus specifically on security in 6G networks, they do mention a mechanism for security functions.
In \cite{chatzistefanidis2026} a tripartite governance model that divides power between three specialized agent branches is used. This division, alongside the use of trust scores for individual agents, ensures the secure and fair execution of intents. Prior to the execution, the execution agent verifies the current state of the network in real time to enforce the ratified intents securely. This is done by integrating intents with network telemetry to send precise, verified resource instructions to the 6G radios.
The framework in \cite{Brodimas20257150} outlines how intent security is achieved by classifying the tools agents use. For critical decisions or when a sensitive tool is used that could interfere with the infrastructure, an alert is sent to a human-in-the-loop to confirm the agent's decision. 
Paper \cite{Chatzistefanidis2024227} relies on an agent they refer to as the Arbitral Unit. This agent analyses the other agents' reasoning during resource negotiations. If it recognizes a motive as malicious, it activates incentives such as warnings or penalties. 
None of these three frameworks has a dedicated security agent. They use different concepts to keep their systems operable, but they do not explicitly refer to network security; rather, they protect themselves from overload caused by incorrectly configured intents or greedy agents. Our framework differs in that we have agents that deal exclusively with network security.
\begin{table*}[bth]
  \vspace{0.5\in}
  \caption{Comparison of agentic \ac{IBN} approaches with focus on security in telecommunication networks}  
  \label{tab:comparison}
  \centering
   \begin{tabular}{ccclcc}
    \toprule
    \textbf{Related work}           & \textbf{Agentic IBN} &   \textbf{Secure IBN}& \textbf{Application} & \textbf{Multi-Domain} &\textbf{Agentic Workflow Topology} \\
    %                               &                      &                      &                      &                       &\textbf{ Topology} \\
    \midrule
    \cite{chatzistefanidis2026}     & \checkmark  &(\checkmark )&  6G RAN                 & \checkmark     & Hierarchical \\
                                    &             &             & Automation/Marketplace  &                &              \\
    %\cite{tu2025intent}            &      x      &      x      &  Network Configuration  & x              & LLM pipeline\\
    % \cite{jiang2026agentic}         & \checkmark  &      x      &  Network Slicing        & \checkmark     & Hierarchical\\
    \cite{Mekrache202648}           & \checkmark  &      x      & Cross Domain 6G         &  \checkmark    & Hierarchical\\
                                    &             &             & OSS Management          &                &             \\
     \cite{Mekrache2025158}         & \checkmark  &      x      &6G OSS \&                & \checkmark     & Hierarchical\\
                                    &             &             & Service Lifecycle       &                &             \\
     \cite{10.1145/3718958.3750537} & \checkmark  &      x      & Enterprise/Datacenter   & x              & Hierarchical\\
                                    &             &             & Management              &                &             \\
     \cite{Brodimas20257150}        & \checkmark  &(\checkmark )& Cloud-Native            & \checkmark     & Hierarchical\\
                                    &             &             & Infrastructure (K8s)     &                &              \\
     \cite{MartinezJulia2025}       & \checkmark  &      x      &Intent Resolution \&     & \checkmark     & Collaborative\\ 
                                    &             &             & Mapping                 &                &           \\
     \cite{Avgerinos2025}           & \checkmark  &      x      & Fault Management \&     & \checkmark     & Closed Loop\\ 
     \cite{MartinezJulia20251}      &\checkmark   &      x      & Intent Mapping \&       & x              & Closed Loop\\ 
                                    &             &             & Resolution              &                &              \\
     \cite{Li202512}                & \checkmark  &      x      & (Com)$^2$ Net Config    & \checkmark     & Hierarchical \\ 
     \cite{Chatzistefanidis2024227} & \checkmark  &(\checkmark )& SLA Negotiation \&      & \checkmark     & Collaborative \\ 
                                    &             &             & Throughput              &                &               \\
    \cite{Araujo2024}               & \checkmark  &      x      & SDN Traffic             & x              & Closed Loop \\ 
                                    &             &             & Engineering             &                &             \\
    \textbf{Our work}               & \textbf{\checkmark}  & \textbf{\checkmark}  &  \textbf{Security in 6G }     & \textbf{\checkmark}     & \textbf{Horizontal} \\
    \bottomrule
  \end{tabular}
 \end{table*}

%Subgroup 2 no security
%subgroup 2a no security and hierarchical workflow
The focus of the second group of papers surveyed is on networking, ensuring a running system, rather than on secure IBN. The group is split into three subgroups, each of which is based on the agentic workflow architecture used by the group. Due to the novel nature of the topic, there is currently no standardized, uniform definition of topology types. The various frameworks surveyed presented hierarchical, collaborative, and closed-looped topologies, which we adopted as categories for this survey. In hierarchical topologies, agency is distributed through top-down delegation, in which orchestrator agents decompose complex goals into sub-tasks for subordinate agents. In contrast, collaborative topologies facilitate horizontal agency. Agents engage in bidirectional communication and often use peer-review mechanisms to refine collective outputs iteratively. At last, closed-loop systems make use of ongoing environmental or programmed feedback for self-correction. These systems can take the form of horizontal or collaborative configurations, in which the evaluation process is distributed across a network of agents rather than being overseen by a single supervisor. Through this recursive peer evaluation, the system iteratively refines its output until it meets the defined success criteria.
% The frameworks in the first subgroup utilize a hierarchical organization of agents. 
%In this workflow, tasks are delegated down the hierarchy. Higher-level agents dividing jobs and delegate them to the lower-level agents.

In \cite{Mekrache2025158}, OSS-GPT is introduced as a foundational intelligence layer that uses LLMs to automate Operation Support System (OSS) tasks by translating natural language into network management commands. This logic is then integrated in \cite{Mekrache202648}, which scales the architecture into a distributed management and orchestration framework where LLM-driven entities operate across multiple domains to resolve intents.
The multi-agent framework Confucius, presented in paper \cite{10.1145/3718958.3750537} uses LLM reasoning to couple domain-specific tools with existing foundation models. It translates natural language intents into executable configurations while maintaining a continuous closed loop to ensure the network remains in the desired state.
Unlike the other frameworks referenced, \cite{Li202512} utilizes a centralized, single-agent architecture that leverages an LLM to translate between a Human operator and the complex $(Com)^2Net$ Infrastructures. 
The collaborative multi-agent tool presented in \cite{MartinezJulia2025} is used for intent resolution by organizing distributed agents into a peer-to-peer chain via a simplified Chord algorithm. This allows for sequential distributed AI inference across multiple domains, where each agent contributes a shared Knowledge Object until a complete network service definition is achieved.
% Not peer-revied
% The framework introduced in \cite{jiang2026agentic} has an agent similar architecture to the previous paper. \hl{ NOT CLEAR It uses a hierarchical multi-agent system that integrates through an Orchestrator Agent that coordinates domain-specific specialist }agents, for RAN and Core networks using ReACT (Reasoning and Acting) cycles. 
%subgroup 2b no secure IBN- Collaborative workflow
%The second subgroup deals with multi-agent frameworks, where the agentic workflow is collaborative. \hl{ups the previous ones are NOT collaboraive??? how come???}%Collaboration is achieved through multiple agents, either assigning tasks to themselves or having tasks assigned to them by a lead orchestrator. The agents communicate with each other, peer-reviewing and refining each other's outputs in a recursive loop until a solution is found. 
%subgroup 2c no security and cloosed loop
% A closed-loop approach is used by the last group of frameworks. 
%In these systems, the output of the agents is continuously monitored by a feedback loop and evaluated against a set of requirements. The output is refined until it matches the objective success criteria set by the user or environment.

A closed-loop agentic framework that introduces a self-healing mechanism for cloud, edge, and IoT networks is presented in \cite{Avgerinos2025}. This mechanism uses root cause analysis to trigger fault mitigation. The Multi-Agent Framework in \cite{MartinezJulia20251} operates within a single-domain intent resolution and demonstrates how decentralized agents can translate natural-language intent into network-service descriptions. 
Similar to the current work, \cite{Araujo2024} employs \acp{LLM} and a multi-agent architecture to control a mobile network using high-level intents. 
However, since the architecture presented in this subgroup adopts a closed-loop, sequential approach, extending the \ac{IBN} platform through external third-party agents is very difficult due to the agents' flat organization. 

% \hl{where is this gap visible from the table?} 
This paper contributes to the body of work by proposing an intent-based security architecture that addresses a crucial gap in enabling autonomous functioning in multi-agent and multi-domain networks by providing a robust interface between the network and its operator. Our work mainly focuses on the adaptation of the agentic AI paradigm to the existing \ac{IBN} architecture \cite{rfc9315}, while retaining the essential functional building blocks. The proposal primarily enhances network security while also possessing the adaptability to be used for other applications, as described in ~\cite{irtf-ibn-usecases}. 

% \hl{not clear the taxonomy. seemslike hierarchical is not collaborative, closed loop is.,...not clear too many concepts that sound similar}

%In other words, the agents are organized in a flat structure, making it difficult to extend the \ac{IBN} platform through external or third-party agents.

% Finished: 2026-02-02 18:50 -- Iuli
\section{A novel multi-agent intent-based security framework}
\label{sec:architecture}

% \acrodef{OA}{orchestrator agent}

% \begin{mindflow}
%     Placeholder for the new architecture, followed by the analysis of why it is better, and covers the requirements.
% \end{mindflworkflowow}

 This section provides the technical details of the proposed intent-based security architecture, the workflow within the architecture, and an example of the intent pipeline. The architecture shown is designed for network security as it is handled in the EU Project Mare, but is generic enough to be used in other 6G architectures. We assume that the network operator is the network manager or a person in charge of configuring and maintaining the network (\ie network manager role).
 
\subsection {The architecture}
\begin{figure*}[tb]
  \centering
\includegraphics[width=.70\linewidth]{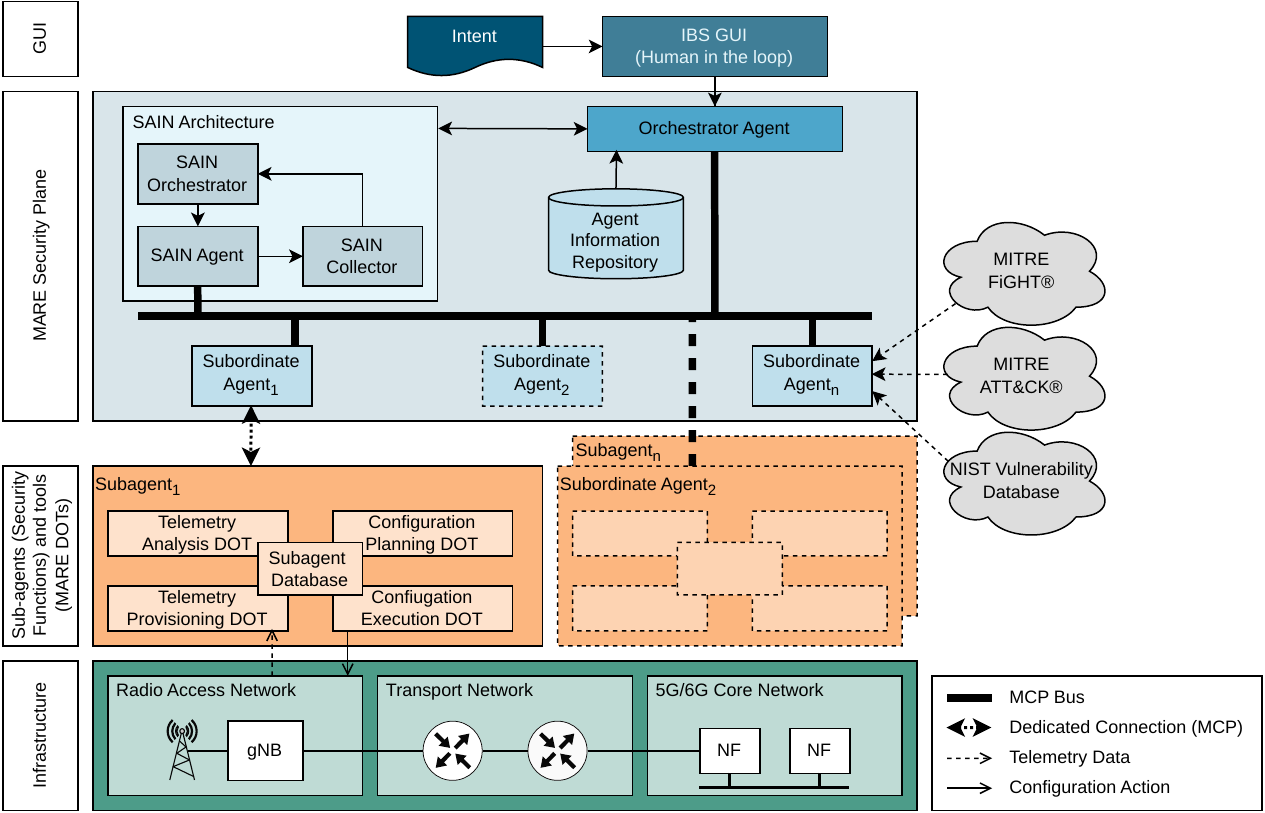}
  \caption{High-level representation of a multi-agent intent-based 
  security framework and its connection with the infrastructure 
  and external data repositories, including SAIN architecture from RFC 9417~\cite{Claise2023SAIN}.}
  \label{fig:architecture}
  %\todo{(1) align the line connecting Subordinate Agent 1; (2) change names of Subagent to Sub-agent}
\end{figure*}

The architecture depicted in \cref{fig:architecture} is split into four layers: \ac{GUI}, MARE Security Plane, Sub-agents and tools, and Infrastructure Layer. The GUI layer contains the components that interface with the user (in our case, the network manager), allowing the user to enter the intent in natural or controlled language. The \ac{GUI} connects to the \ac{OA} via an HTTP request using the REST model, allowing external tools to access the same API endpoint. 

% -----------------------------------------------------------------
% -----------------------------------------------------------------

The MARE Security Plane accommodates the \ac{OA}, \ac{AIR}, Subordinate Agents, and the components that compose the SAIN Architecture, proposed in \cite{Claise2023SAIN}. The \ac{OA} is the main block of this architecture, translating intents specified in a high-level specification into actions that can be applied to the network, following an approach similar to~\cite{dzeparoska2024}. The \ac{AIR} is a database containing information about the network topology, network security functions, available agents and tools, how to call the agents and tools (\eg the function signature), and historical data about previous decompositions. Additionally, this repository may also maintain a dictionary of known attacks and mitigation procedures. The $\text{Subordinate Agents}_{n}$, where $n \in N = \{1,\cdots, N\}$, is a collection os agents that receive commands from the \ac{OA} and are in charge of deploying the actions (\eg, changing configurations or deploying new functions) in the network withing the domain they can control. The Subordinate Agents are not necessarily placed at the MARE Security Plane. Domains external to the MARE Security Plane can host their agents and expose an interface to the \ac{OA}. The agents communicate with the \ac{OA} via a common, IP-based bus, similar to the bus used to connect \acp{NF} in the 5G core. Messages are encoded using the \ac{MCP} and transported via HTTP streams, allowing bidirectional communication. The bus is illustrated by the dashed bold black line in \cref{fig:architecture}. Those agents can act within their domain using tools or by delegating tasks to Sub-agents. Subordinate Agents can also perform tasks that are not strictly related to network control. In \cref{fig:architecture}, $\text{Subordinate Agent}_n$ is used to communicate with external data sources. Finally, the blocks proposed in the \ac{SAIN} Architecture can optionally be integrated into the proposed MARE Security Plane, providing service assurance functions as described in~\cite{Claise2023SAIN}.

The Sub-agents and tools layer, at the center of \cref{fig:architecture}, comprises a set of tools that can provide network information, deploy new \acp{NF}, and reconfigure the network within the domain they are placed, or optionally, be organized as more elaborate Security Functions that are able to make low-level decisions. In both cases, they are controlled by their respective Subordinate Agent, that communicate with the tool via a dedicated \ac{MCP} connection. The communication between the tools and the infrastructure should follow the interface exposed by the network element to be controlled. This communication can be done using automation tools and frameworks (such as pyATS\footnote{https://developer.cisco.com/pyats/} and NAPALM\footnote{https://github.com/napalm-automation/napalm} to provide vendor-agnostic interfaces to the network elements.

At the bottom of \cref{fig:architecture}, the infrastructure layer comprises the network elements controlled by agents and tools. These elements can be physical or virtual network functions, as well as the underlying infrastructure that supports them. Due to the heterogeneity of the infrastructure, it is split into different domains, each of them having its own set of tools and subordinate agents.

% \hl{restructure this section. in the first part describe the figure. like "it consists of GUI, MARE SUbagent and INfra, and describe EVERY Box underneath. In the subsequent subsection, descibe the workflow how it works together...do not mix the two make sure that every box and abbreviation is explained}

% Removed
% TODO: Check if this information is already above 
% Data from the Agent Information Repository is fetched via \ac{RAG} techniques to provide the \ac{LLM} running in the \ac{OA} with updated 

% This repository might also be updated by agents consuming external data from vulnerability databases (\eg, MITRE ATT\&CK, MITRE FiGHT and NIST), thus allowing the system to react to 0-day threats. 

% Also, it is important to notice that the agents do not necessarily need to be in the same administration domain as the \ac{OA}. For example, Subordinate Agent$_2$ could be an external agent, illustrated as the dashed orange block at the right of \cref{fig:architecture}. Those agents can also be provided by a third-party developer, allowing external agents to be plugged in later, provided that their interfaces and function call signatures are available at the Agent Information Repository. Finally, Service Assurance is performed according to the \ac{SAIN} Architecture~\cite{Claise2023SAIN}. However, due to the specifics of our proposed architecture, the Orchestrator Agent will also perform the tasks that are the responsibility of the \ac{SAIN} Orchestrator.

\subsection{The workflow}

The system's input is a high-level security intent expressed in natural or controlled language entered by the network manager. As illustrated at the top of \cref{fig:architecture}. The intent is entered using a \ac{GUI} from the \ac{IBS}, or via an external application via REST \ac{API}. The \ac{OA} receives the intent and processes it, transforming the high-level intent into tasks and actions that can be executed by the tools connected to the network elements. 

\begin{figure}[tb]
  \centering
  \includegraphics[width=0.75\linewidth]{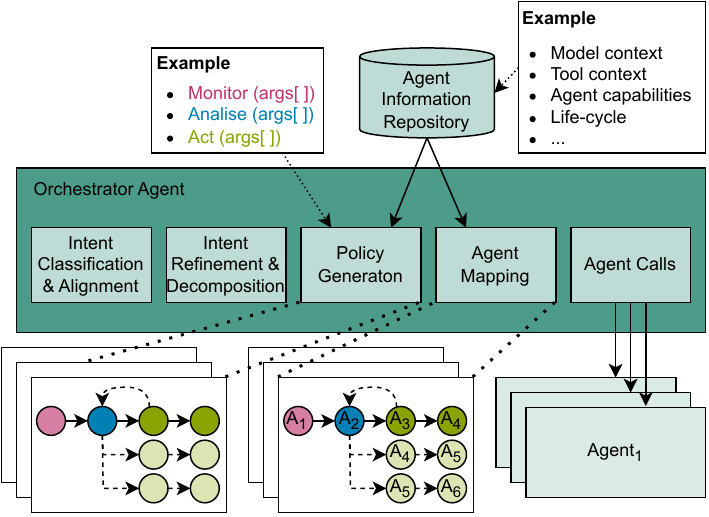}
  \caption{Intent pipeline processing, carried out by the Orchestrator Agent.}
  \label{fig:orchestrator}
\end{figure}

An intent pipeline that runs within the \ac{OA} is illustrated in \cref{fig:orchestrator}. After receiving the intent, the \ac{OA} classifies it and verifies whether the \ac{IBS} can process it. When the \ac{IBS} cannot process the intent, the network manager receives an alert, and the intent processing finishes. When the \ac{IBS} can handle the intent, it starts the alignment process, which transforms the intent into a formalized representation, using a predefined data structure. As in~\cite {dzeparoska2024}, \acp{LLM} are used in this step. The alignment process identifies whether required information is missing and employs a human-in-the-loop process to request the network manager to provide the missing parameters. The alignment process is iterative and might require multiple executions to complete. The result of this phase is a structured representation of the intent, its type, time information and additional metadata that help the \ac{LLM} understand the user's intent. The formalization of the intent aims to eliminate intent misalignment and possible ambiguity on the intent. 

% -----------------------------------------------------------------
% -----------------------------------------------------------------

In the next step, the Intent Refinement phase starts. Similar to \cite{dzeparoska2024}, the Intent Refinement and Decomposition block receives the formalized intent (\textit{declarative intent}) and decomposes it into lower-level, more specific (\ie \textit{definitive} and \textit{imperative}) policies. The two sets of policies specify what should be done in the network to achieve the goals set by the intent. The policies are then sent to the Policy Generation block, which maps them to a feasible set of actions represented as a graph. \textit{Definitive} policies are mapped to monitoring, analysis, or low-level planning policies, while \textit{imperative} policies are basically mapped to actions or configuration changes~\cite{dzeparoska2024}. The policy generation process is guided by data stored in the \ac{AIR}, enabling the \ac{IBS} to determine the set of available actions. Instead of the progressive policy generation/execution loop proposed in~\cite{dzeparoska2024}, our proposed architecture generates multiple policy sets that may fulfill the intent. This process avoids calling the policy generation for every execution iteration. If one of the policy sets fails, the executed configuration changes are rolled back~\cite{chen-undo-2025}, restoring the system to its previous state, and then a new set of policies is executed. Next, the Agent Mapping maps each low-level policy to agents capable of executing the corresponding actions, as shown at the bottom of \cref{fig:orchestrator}. This process is supported by data from the \ac{AIR}, and the system tries to assign the agents that better fit the task. If a policy or configuration action cannot be assigned to an agent, the whole set of actions is discarded, and the Agent Mapping will proceed with the next available set of policies. 

Finally, the last step of the workflow is the Progressive Agent Calls, which coordinates and routes each function call to the agents assigned in the previous step. Each agent is called sequentially, following the predefined order. The agents report the result of the action to the Progressive Agent Calls block. If a policy or configuration change cannot be deployed on the network, the Progressive Agent Calls block coordinates with the agents to roll back the previously applied configuration and proceed with the next set of actions. Following the example shown in \cref{fig:orchestrator}, the agent $A_{1}$ is initially called to start a monitoring task that reports data to the \ac{IBS}. Next, agent $A_{2}$ is called to analyze the monitoring data and propose modifications to the network when required. Agent $A_{3}$ is called to perform corrective action when required. If the agent $A_{3}$ fails to perform the assigned action, it informs the Orchestrator Agent, which selects an alternative path for execution, in this example, either calling $A_{4} \rightarrow A_{5}$ or $A_{5} \rightarrow A_{6}$.
\section{Verification}
\label{sec:verification}

There are no specific standards today for evaluating the IBS system, but various options are available. In our approach, we develop a functional multi-agent \ac{IBS} prototype and provide initial results; however, further verification methods are needed, such as assessing whether the intent was well interpreted or the LLM was well trained. In our implementation, the \ac{OA} and Subordinated Agents were developed using the LangChain\footnote{https://github.com/langchain-ai/langchain} and LangGraph\footnote{https://github.com/langchain-ai/langgraph} Python libraries. The prototype implements the core modules described in \cref{fig:architecture}, except the SAIN components. The Orchestrator Agent employs \ac{LLM} models to perform intent classification, alignment, refinement, and decomposition. We conducted tests on four models of varying sizes, all of which had reasoning capabilities. \cref{tab:models} list the tested \acp{LLM}. Three of the tested models are classified as \textit{small} models (OpenAI GPT o4-mini, Mistral AI Magistral Small 2509, Qwen3 30B-A3B), with the last three models 

running on-premises.

\begin{table}
  \caption{Models used to implement the Orchestrator Agent}
  \label{tab:models}
  \centering
  \begin{tabular}{lll}
    \toprule
    \textbf{Model Name}             & \textbf{Estimated}  & \textbf{On-Premise} \\
                                    & \textbf{Parameters} &                     \\
    \midrule
    OpenAI o4-mini                  & 8B -- 40B  & No     \\
    Mistral AI Magistral Small 2509 & 24B        & Yes    \\
    OpenAI GPT-OSS 120b             & 120B       & Yes    \\
    Qwen Qwen3 30B-A3B              & 30.5B      & Yes    \\ 
    \bottomrule
  \end{tabular}
\end{table}

% done!
To allow the Orchestrator Agent to perform task classification and later task routing, we fed it with data about the available subordinate agents, including the tools and tasks each subordinate agent can perform within the domain where the agent operates. The orchestrator agent also implements an intelligent dispatcher that routes calls to the competent agents. Actions are sent to subordinate agents as a function with the corresponding parameters, which can be encoded and transmitted via \ac{MCP}. Since we are interested in the overall system performance, subordinate agents do not implement configuration changes on the network devices. Instead, agents log their actions, allowing us to assess whether the action was correctly routed to the agent and whether the actions correctly arrived at the agent.

\begin{figure}[tb]
  \centering
\includegraphics[width=1.0\linewidth]{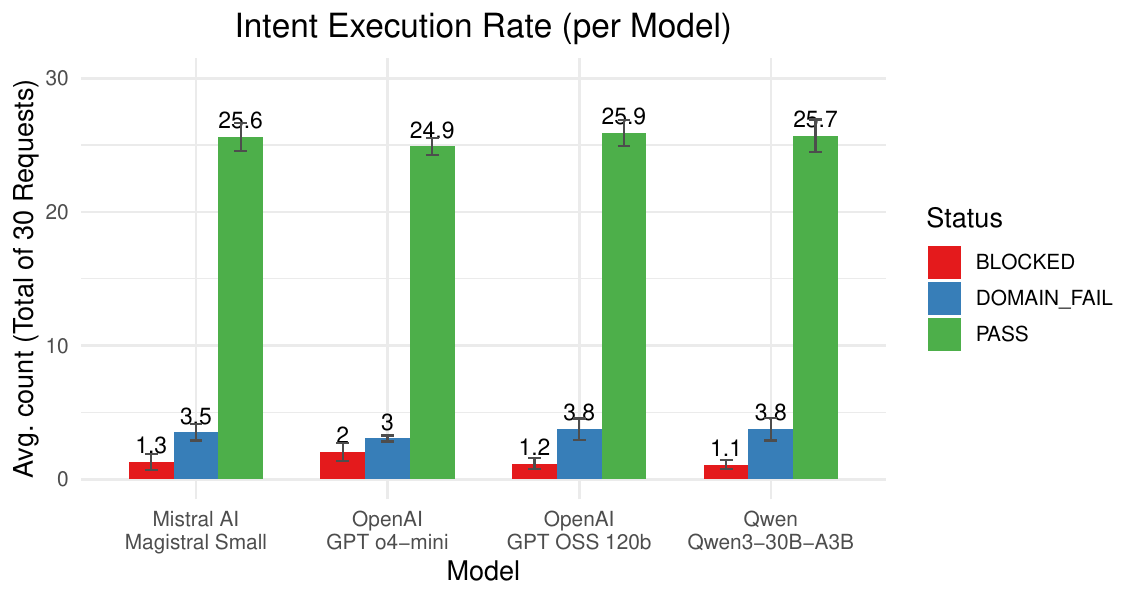}
  \caption{Success rate of intent execution with the proposed architecture}
  \label{fig:results}
\end{figure}

% Use Case
For the evaluation of the system, we created three sets of user intent focusing on the security of the mobile networks. \Cref{lst:intent-sample} presents some examples of crafted intents. Each set contains 10 different intents. Each set was progressively less descriptive than the previous one. Each prompt has been designed to address a particular network configuration, and each configuration must be implemented in a specific section of the network. Our experiment was designed to evaluate the whole system, and the produced results are categorized into three categories: pass, domain fail, and blocked. Pass indicates that the Orchestrator agent correctly understands the intent, decomposes it into smaller \textit{definitive} intents, derives configuration actions, correctly routes the actions to the subordinate agent, and the agent receives and prints the action to the standard log. A domain fail indicates correct parsing and processing of the intent, but wrong routing of the configuration actions, not arriving at the expected domain. Finally, blocked indicates that the \ac{IBS} could not decompose the intent into actions because the LLM does not find a match between the requested intent via prompt and the known configuration actions. 
% -----------------------------------------------------------------
\begin{lstlisting}[
  float=bth,
  caption={Example of intent given to the orchestrator agent},
  label={lst:intent-sample},
  language={},
  basicstyle=\ttfamily\footnotesize,
  backgroundcolor=\color{lstbg},
  frame=single,
  rulecolor=\color{lstframe},
  frameround=tttt,
  framesep=4pt,
  xleftmargin=1.2em,
  xrightmargin=1.2em,
  aboveskip=0.6\baselineskip,
  belowskip=0.2\baselineskip,
  breaklines=true,
  breakatwhitespace=true,
  columns=fullflexible,
  keepspaces=true,
  showstringspaces=false,
  numbers=none,
  lineskip=1pt,
]
- Ensure the Access and Mobility Management Function remains resilient against signaling-based exhaustion.
- Guarantee that permanent subscriber identities are never transmitted in an unencrypted state.
- Maintain a network environment free from unauthorized or malicious base station connections. 
- Protect all user-plane traffic transitioning between the radio access and core networks from interception.
\end{lstlisting}
% -----------------------------------------------------------------
\Cref{fig:results} presents the results of the carried out experiment. We performed 20 iterations of the experiment, submitting the entire set of intents to the \ac{IBS}. During our tests, we noticed that one intent prompt and its variants caused the \ac{IBS} to fail, resulting in a blocked outcome. While a deep investigation is required to precisely pinpoint the reason why the intent ``Strictly limit the use of regulatory interception functions to verified lawful requests'' causes the system to fail, we conjecture two possible reasons. Since no subordinate agent is mapped to any function related to lawful interceptions, the system might not be able to find a suitable agent, and the intent could not be mapped to a domain. The second hypothesis is that, given the semantics of the intent prompt, which is strictly linked to legal and regulatory frameworks, the \ac{LLM} halts processing due to a lack of knowledge. While the first hypothesis can be easily fixed by adding a new agent or by tuning the context description of a capable subordinate agent to better relate to this type of intent, the second hypothesis might require a more elaborate approach, similar to what is described in~\cite{chatzistefanidis2026}.

% -----------------------------------------------------------------
% -----------------------------------------------------------------

All tested models produced acceptable results, even those classified as \textit{small}. This is an important characteristic of the proposed architecture because it allows network operators and small \acp{ISP} to deploy the system on their premises, thereby partially fulfilling the security and privacy requirements. The agentic and modular approach adopted by our proposed architecture enables the integration of blocks and functions from the \ac{SAIN} architecture, thus covering telemetry and assurance requirements. Finally, we adopt an agentic \ac{IBN} architecture by design, allowing the \ac{IBS} to learn and expand its knowledge base using external sources. 
\section{Conclusions and Outlook}
\label{sec:conclusions}

This paper presents a novel \ac{IBN} framework designed to manage security mechanisms in 6G networks by integrating agentic AI. By proposing an \ac{IBN} framework to address that integrates with the Security Plane proposed in the scope of the EU project MARE, we studied how the complexity of next-generation infrastructure can be managed by using high-level security goals rather than low-level configurations, allowing the network operation to focus in \textit{what} should be achieved, instead of focusing on the how network devices are configured.

We implemented a novel hierarchical multi-agent architecture that utilizes an interactive intent-processing pipeline. Our architecture bridges the gap between natural-language intents and actionable network policies and configurations by employing \acp{LLM}. Furthermore, we demonstrated its reasoning capabilities by enriching its context with external knowledge bases, such as MITRE and NIST, which are critical for defending against zero-day threats. Experimental verification of our prototype confirmed that even small on-premises LLM models can achieve high intent-execution success rates. This indicates that the proposed architecture is viable for deployment by various network operators, ensuring performance and privacy by avoiding sending sensitive network data to third-party companies executing the \ac{LLM}.

Future work should address the implementation and integration of the functions described in the IETF SAIN architecture while keeping compatibility with the agentic approach. Development and deployment of subordinate and sub-agents interfacing with a full 5G network could yield interesting insights into the probability of success of agents when performing their tasks. 
%%% ===============================================================================
%%% Acknowledgment
%%% ===============================================================================
\section*{Acknowledgment}

This work was performed in the context of the MARE project, which has received funding from the Smart Networks and Services Joint Undertaking (SNS JU) under the European Union’s Horizon Europe research and innovation programme under Grant Agreement No 101191436.

%%% ===============================================================================
%%% Bibliography
%%% ===============================================================================
\bibliographystyle{IEEEtranN} % IEEEtranN is the natbib compatible bst file
\bibliography{paper}

@misc{rfc9315,
  series       = {Request for Comments},
  number       = 9315,
  howpublished = {RFC 9315},
  publisher    = {RFC Editor},
  doi          = {10.17487/RFC9315},
  author       = {Alexander Clemm and Laurent Ciavaglia and Lisandro Zambenedetti Granville and Jeff Tantsura},
  title        = {{Intent-Based Networking - Concepts and Definitions}},
  pagetotal    = 23,
  year         = 2022,
  month        = oct,
}

@article{dzeparoska2024,
  author={Dzeparoska, Kristina and Tizghadam, Ali and Leon-Garcia, Alberto},
  journal={IEEE Comm. Magazine}, 
  title={Emergence: An Intent Fulfillment System}, 
  year={2024},
  volume={62},
  number={6},
  pages={36-41},
  doi={10.1109/MCOM.001.2300270}
}

@techreport{gu2025,
  number =    {draft-gu-nmrg-intent-translator-02},
  type =      {Internet-Draft},
  institution =   {Internet Engineering Task Force},
  publisher = {Internet Engineering Task Force},
  note =      {{Work in Progress}},
  OPTurl =       {https://datatracker.ietf.org/doc/draft-gu-nmrg-intent-translator/02/},
  author =    {Mose Gu and Jaehoon Paul Jeong},
  title =     {{An Intent Translation Framework for Internet of Things}},
  pagetotal = 17,
  year =      2025,
  month =     oct,
  day =       20,
}

@inproceedings{chen-undo-2025,
    author={Yinfang Chen and Jiaqi Pan and Jackson Clark and Yiming Su and Noah Zheutlin and Bhavya Bhavya and Rohan Arora and Yu Deng and Saurabh Jha and Tianyin Xu},
    booktitle = {Proc. of the Annual Conf. on Neural Information Processing Systems},
    title={STRATUS: A Multi-agent System for Autonomous Reliability Engineering of Modern Clouds}, 
    year={2025},
    doi = {10.48550/arXiv.2506.02009},
}

@article{xiao2025,
  author={Xiao, Yong and Shi, Guangming and Zhang, Ping},
  journal={IEEE Comm. Magazine}, 
  title={Toward Agentic AI Networking in 6G: A Generative Foundation Model-as-Agent Approach}, 
  year={2025},
  volume={63},
  number={9},
  pages={68-69},
  doi={10.1109/MCOM.001.2500005}
}

@article{chatzistefanidis2026,
  title = {{AGORAN}: An agentic open marketplace for {6G} {RAN} automation},
  journal = {Computer Networks},
  volume = {275},
  pages = {111927},
  year = {2026},
  issn = {1389-1286},
  doi = {https://doi.org/10.1016/j.comnet.2025.111927},
  OPTurl = {https://www.sciencedirect.com/science/article/pii/S138912862500893X},
  author = {Ilias Chatzistefanidis and Navid Nikaein and Andrea Leone and Ali Maatouk and Leandros Tassiulas and Roberto Morabito and Ioannis Pitsiorlas and Marios Kountouris}
}

@techreport{irtf-ibn-usecases,
  number =    {draft-irtf-nmrg-ibn-usecases-02},
  type =      {Internet-Draft},
  institution = {Internet Engineering Task Force},
  publisher = {Internet Engineering Task Force},
  note =      {Work in Progress},
  url =       {https://datatracker.ietf.org/doc/draft-irtf-nmrg-ibn-usecases/02/},
  author =    {Kehan Yao and Danyang Chen and Jaehoon Paul Jeong and Qin Wu and Chungang Yang and Luis M. Contreras and Giuseppe Fioccola},
  title =     {{Use Cases and Practices for Intent-Based Networking}},
  pagetotal = 49,
  year =      2025,
  month =     nov,
  day =       3,
}

@Manual{mwe,
  title  = {The \texttt{mwe} Package},
  author = {Martin Scharrer},
  year   = {2017},
  url    = {http://texdoc.net/mwe},
}

@inproceedings{tong2025core,
  author={Tong, Wen and Huo, Wei and Lejkin, Thierry and Penhoat, Joel and Peng, Chenghui and Pereira, Carlos and Wang, Fei and Wu, Shaoyun and Yang, Lu and Shi, Yuanming},
  booktitle={IEEE Int. Conf. on Comm. Workshops (ICC Workshops)}, 
  title={A-Core: A Novel Framework of Agentic AI in the 6G Core Network}, 
  year={2025},
  volume={},
  number={},
  pages={1104-1109},
  doi={10.1109/ICCWorkshops67674.2025.11162291}
}

@article{antonopoulos2025agile,
  author={Antonopoulos, Angelos and Kartsakli, Elli and Bartzoudis, Nikolaos and Brodimas, Dimitrios and Vukobratovic, Dejan and Tsolkas, Dimitris and Diego, Ferran and Roumelas, George and Tomaszewski, Lechosław},
  journal={IEEE Network}, 
  title={AGILE-6G: Agentic AI for Autonomous Management of 6G Network/Application Services}, 
  year={2025},
  volume={},
  number={},
  pages={1-9},
  doi={10.1109/MNET.2025.3644039}
}

@article{zhang2026toward,
  author={Zhang, Ruichen and Liu, Guangyuan and Liu, Yinqiu and Zhao, Changyuan and Wang, Jiacheng and Xu, Yunting and Niyato, Dusit and Kang, Jiawen and Li, Yonghui and Mao, Shiwen and Sun, Sumei and Shen, Xuemin and Kim, Dong In},
  journal={IEEE Comm. Surveys \& Tutorials}, 
  title={Toward Edge General Intelligence With Agentic AI and Agentification: Concepts, Technologies, and Future Directions}, 
  year={2026},
  volume={28},
  number={},
  pages={4285-4318},
  doi={10.1109/COMST.2026.3651702}
}

@INPROCEEDINGS{SecurityIBNAhmad2023,
  author={Ahmad, Ijaz and Malinen, Jere and Christou, Filippos and Porambage, Pawani and Kirstädter, Andreas and Suomalainen, Jani},
  booktitle={IEEE Conf. on Standards for Comm. and Networking (CSCN)}, 
  title={Security in Intent-Based Networking: Challenges and Solutions}, 
  year={2023},
  volume={},
  number={},
  pages={296-301},
  doi={10.1109/CSCN60443.2023.10453125}}

@INPROCEEDINGS{Pillai2024SDNDDoS,
  author={Vadakkethil Somanathan Pillai, Sanjaikanth E and Polimetla, Kiran},
  booktitle={Int. Conf. on Integrated Circuits and Comm. Systems (ICICACS)}, 
  title={Mitigating {DDoS} Attacks using {SDN-based} Network Security Measures}, 
  year={2024},
  volume={},
  number={},
  pages={1-7},
  doi={10.1109/ICICACS60521.2024.10498932}}

@article{leivadeas2022survey,
  title={A survey on intent-based networking},
  author={Leivadeas, Aris and Falkner, Matthias},
  journal={IEEE Comm. Surveys \& Tutorials},
  volume={25},
  number={1},
  pages={625--655},
  year={2022},
  publisher={IEEE}
}

@techreport{Ahn2025I2NSF5G,
  author      = {Yonghwan Ahn and Jaehoon Jeong and Yongseok Kim},
  title       = {An Integrated Security Service System for {5G} Networks using an {I2NSF} Framework},
  institution = {IETF Internet-Draft},
  type        = {Work in Progress},
  number      = {draft-ahn-opsawg-5g-security-i2nsf-framework-00},
  year        = {2025},
  month       = {July}
}

@article{tu2025intent,
  title = {Intent-Based Network Configuration Using Large Language Models},
  author = {Tu, Nguyen and Nam, Sukhyun and Hong, James Won-Ki},
  journal = {Int. Jour. of Network Management},
  volume={35},
  number={1},
  pages={e2313},
  year={2025},
  publisher={Wiley Online Library},
  doi = {https://doi.org/10.1002/nem.2313},
}

@techreport{Claise2023SAIN,
  author      = {Benoit Claise and Jean Quilbeuf and Diego Lopez and Daniel Voyer and Thyla Arumugam},
  title       = {Service Assurance for Intent-Based Networking Architecture},
  institution = {Internet Engineering Task Force (IETF)},
  type        = {RFC},
  number      = {9417},
  year        = {2023},
  month       = {July},
  note        = {Informational RFC}
}

@article{ribeiro2022deterministic,
  title={A deterministic approach for extracting network security intents},
  author={Ribeiro, Rafael Hengen and Jacobs, Arthur Selle and Zembruzki, Luciano and Parizotto, Ricardo and Scheid, Eder John and Schaeffer-Filho, Alberto Egon and Granville, Lisandro Zambenedetti and Stiller, Burkhard},
  journal={Computer Networks},
  volume={214},
  pages={109109},
  year={2022},
  publisher={Elsevier}
}

@ARTICLE{Mekrache202648,
	author = {Mekrache, Abdelkader and Ksentini, Adlen and Verikoukis, Christos},
	title = {DMO-GPT: An Intent-Driven Framework for Distributed 6G Management and Orchestration},
	year = {2026},
	journal = {IEEE Comm. Magazine},
	volume = {64},
	number = {1},
	pages = {48 – 54},
	doi = {10.1109/MCOM.001.2500258},
}

@CONFERENCE{Mekrache2025158,
	author = {Mekrache, Abdelkader and Ksentini, Adlen and Verikoukis, Christos},
	title = {Next-Generation {6G} Network Management with {OSS-GPT}},
	year = {2025},
	journal = {Proc. of ACM SIGCOMM 2025 Posters and Demos},
	pages = {158 – 160},
	doi = {10.1145/3744969.3748429},
}

@inproceedings{10.1145/3718958.3750537,
author = {Wang, Zhaodong and Lin, Samuel and Yan, Guanqing and Ghorbani, Soudeh and Yu, Minlan and Zhou, Jiawei and Hu, Nathan and Baruah, Lopa and Peters, Sam and Kamath, Srikanth and Yang, Jerry and Zhang, Ying},
title = {Intent-Driven Network Management with Multi-Agent {LLMs}: The Confucius Framework},
year = {2025},
isbn = {9798400715242},
publisher = {Association for Computing Machinery},
OPTaddress = {New York, NY, USA},
OPTurl = {https://doi.org/10.1145/3718958.3750537},
doi = {10.1145/3718958.3750537},
booktitle = {Proc. of the ACM SIGCOMM},
pages = {347–362},
numpages = {16},
location = {S\~{a}o Francisco Convent, Coimbra, Portugal},
series = {SIGCOMM '25}
}

@ARTICLE{Brodimas20257150,
	author = {Brodimas, Dimitrios and Birbas, Alexios and Kapolos, Dimitrios and Denazis, Spyros},
	title = {Intent-Based Infrastructure and Service Orchestration Using Agentic-AI},
	year = {2025},
	journal = {IEEE Open Journal of the Comm. Society},
	volume = {6},
	pages = {7150 – 7168},
	doi = {10.1109/OJCOMS.2025.3600706},
}

@INPROCEEDINGS{MartinezJulia2025,
  author={Martinez-Julia, Pedro and Kafle, Ved P. and Asaeda, Hitoshi},
  booktitle={NOMS 2025-2025 IEEE Network Operations and Management Symposium}, 
  title={EDAIR: An Efficient Distributed AI Agent Architecture for Multi-Domain Intent Resolution}, 
  year={2025},
  volume={},
  number={},
  pages={1-7},
  doi={10.1109/NOMS57970.2025.11073742}}

@ARTICLE{Avgerinos2025,
	author = {Avgerinos, Vasilis and Ramantas, Kostas and Alonso, Luis and Verikoukis, Christos},
	title = {ARM: Autonomous Remediation \& Management with LLM Agents for Intent-Driven Control},
	year = {2025},
	journal = {IEEE Internet of Things Journal},
	doi = {10.1109/JIOT.2025.3648858},
}

@CONFERENCE{MartinezJulia20251,
	author = {Martinez-Julia, Pedro and Kafle, Ved P. and Asaeda, Hitoshi},
	title = {A Distributed AI System for Improving Single-Domain Network Intent Resolution},
	year = {2025},
	journal = {Proc. of Conf. on Innovation in Clouds, Internet and Networks, ICIN 2025},
	pages = {1 – 8},
	doi = {10.1109/ICIN64016.2025.10942871},
}

@ARTICLE{Li202512,
	author = {Li, Qing and Xiong, Yanxu and Li, Zonghang and Ma, Chongxi and Yu, Hongfang and Sun, Gang and Luo, Long and Zhang, Zhaofeng},
	title = {KlonetAI: Automating (Com)2Nets Management With Human Language Intents},
	year = {2025},
	journal = {IEEE Network},
	volume = {39},
	number = {3},
	pages = {12 – 19},
	doi = {10.1109/MNET.2024.3507801},
}

@ARTICLE{Chatzistefanidis2024227,
	author = {Chatzistefanidis, Ilias and Leone, Andrea and Nikaein, Navid},
	title = {Maestro: LLM-Driven Collaborative Automation of Intent-Based 6G Networks},
	year = {2024},
	journal = {IEEE Networking Letters},
	volume = {6},
	number = {4},
	pages = {227 – 231},
	doi = {10.1109/LNET.2024.3503292},
}

@CONFERENCE{Araujo2024,
	author = {Araujo, Aramis Sales and das Mercês, Jefferson Maxmiliano O. and da Silva, Rilbert Lima and de Alencar, Allender Vilar and Passos, Iele Facundo and Meneses, Thiago Fonseca and Sousa, Marcelo Portela and Dias, Michel Coura and Santos, Danilo F.S.},
	title = {An Agentic Approach For Dynamic Software-Defined Network Management Using Large Language Models},
	year = {2024},
	journal = {IEEE Conf. on Network Function Virtualization and Software Defined Networks, NFV-SDN 2024},
	doi = {10.1109/NFV-SDN61811.2024.10807498},
}

\end{document}